\documentclass[aps,prd,twocolumn,superscriptaddress,groupedaddress]{revtex4}

\usepackage{amssymb}
\usepackage{color,graphicx}
\usepackage{amsmath}
\usepackage{amsbsy}
\usepackage{amsthm}
\usepackage{bbm}
\usepackage{bm}
\usepackage{epsfig}
\usepackage{lscape}
\usepackage{float}
\usepackage{braket}
\usepackage{comment}
\usepackage[colorlinks=true,citecolor=blue,linkcolor=red,urlcolor=red]{hyperref}
\usepackage[sort&compress]{natbib}
\usepackage[outdir=./]{epstopdf}

\newcommand{\D}{\text{d}}
\newcommand{\com}[2]{\left[ {#1},{#2} \right]}
\newcommand{\acom}[2]{\left\{ {#1},{#2} \right\}}

\newcommand{\bq}{\begin{equation}}
\newcommand{\eq}{\end{equation}}
\newcommand{\bqali}{\begin{equation}\begin{aligned}}
\newcommand{\eqali}{\end{aligned}\end{equation}}
\newcommand\rC{r_\text{\tiny C}}
\newcommand\omegaC{\omega_\text{\tiny C}}
\newcommand\ain{\hat a_\text{\tiny in}}
\newcommand\tcsl{T_\text{\tiny CSL}}
\newcommand\lcsl{\lambda_\text{\tiny CSL}}
\newcommand\kB{k_\text{\tiny B}}
\newcommand\DNS{\mathcal S(\omega)}

\newcommand\rhocm{\hat{\rho}_{\text{\tiny CM}}}

\newcommand\y{{\bm y}}
\newcommand\Q{{\bm Q}}
\newcommand\x{{\bm x}}
\newcommand\p{{\bm p}}

\newcommand\R{{\bm R}}

\graphicspath{{../Figures/}}

\begin{document}
\author{J.~Nobakht}
\affiliation{Department of Physics, Sharif University of Technology, Azadi avenue, Tehran, Iran}
\author{M.~Carlesso}
\email{matteo.carlesso@ts.infn.it}
\affiliation{Department of Physics, University of Trieste, Strada Costiera 11, 34151 Trieste, Italy}
\affiliation{Istituto Nazionale di Fisica Nucleare, Trieste Section, Via Valerio 2, 34127 Trieste, Italy}
\author{S.~Donadi}
\affiliation{Institut f\"ur Theoretische Physik, Universit\"at Ulm, D-89069, Germany}
\author{M.~Paternostro}
\affiliation{Centre for Theoretical Atomic, Molecular and Optical Physics, School of Mathematics and Physics,
Queen's University Belfast, Belfast BT7 1NN, United Kingdom}
\author{A.~Bassi}
\affiliation{Department of Physics, University of Trieste, Strada Costiera 11, 34151 Trieste, Italy}
\affiliation{Istituto Nazionale di Fisica Nucleare, Trieste Section, Via Valerio 2, 34127 Trieste, Italy}

\title{Unitary unravelling for the Dissipative Continuous Spontaneous Localization model: application to optomechanical experiments}
\date{\today}

\begin{abstract}

The Continuous Spontaneous Localization (CSL) model strives to describe the quantum-to-classical transition from the viewpoint of collapse models. However, its original formulation suffers from a fundamental inconsistency in that it is explicitly energy non-conserving. Fortunately, a dissipative extension to CSL has been recently formulated that solves such energy-divergence problem. We compare the predictions of the dissipative and non-dissipative CSL models when various optomechanical settings are used, and contrast such predictions with available experimental data, thus building the corresponding exclusion plots.
\end{abstract}
\maketitle

\section{Introduction}

Collapse models predict the occurrence of the quantum-to-classical transition in light of an intrinsic dynamical loss of quantum coherence, when the mass and complexity of the system increase~\cite{Ghirardi:1986aa,Ghirardi:1990aa,Bassi:2003ab,Bassi:2013aa,Adler:2009aa}. This is achieved by modifying the standard Schr{\"o}dinger equation with the addition of a {\it non-linear} interaction with an external {classical noise field}. The latter induces the localization of the wave function in space. Such interaction is negligible for microscopic systems, and is amplified by an intrinsic in-built mechanism that makes it stronger for macroscopic objects. In this way collapse models account for the quantum behaviour of microscopic systems, as well as for the emergence of classicality in the macroscopic world.

The most studied collapse model is the Continuous Spontaneous Localization (CSL) model \cite{Ghirardi:1990aa}. Here, the interaction of a quantum system with the collapse noise depends on two phenomenological parameters: the collapse rate $\lcsl$, which measures the strength of the noise, and the correlation distance $\rC$, which sets the spatial resolution of the collapse, i.e.~the typical distances above which superpositions are suppressed. 
The quantitative determination of such parameters has been the focus of speculations. The original estimates put forward in Ref.~\cite{Ghirardi:1986aa} have set $\rC=10^{-7}\, $m and $\lcsl=10^{-16}\,\text{s}^{-1}$, later modified to $\lcsl=10^{-9}\,\text{s}^{-1}$, based on the analysis of the process of latent image formation~\cite{Adler:2007ab,Adler:2007ac}.

Recently, a significant amount of work has been devoted to the identification of experiment-based upper bounds on the CSL parameters. Experiments using matter-wave interferometry~\cite{Arndt:1999aa,Eibenberger:2013aa,Toros:2016aa,Toros:2016ab}, entangled macroscopic diamonds \cite{Belli:2016aa}, cantilevers \cite{Vinante:2016aa,Vinante:2016ab}, cold atoms \cite{Laloe:2014ab,Bilardello:2016aa}, X-rays emission~\cite{Curceanu:2016aa,Piscicchia:2017aa}, and gravitational wave detectors \cite{Carlesso:2016ac,Helou:2016aa} have been instrumental to the drawing of an {\it exclusion plot} aiming at narrowing down the range of acceptable values for the collapse parameters.  

A well-known drawback of the phenomenological nature of the CSL model is the prediction of a constant increase of the kinetic energy of a system due to its interaction with the collapse noise. The most conservative prediction of the rate of energy increase is in the range of $10^{-15}$\,K/year~\cite{Bassi:2003aa} (which becomes $10^{-7}$\,K/year when the parameters predicted in Ref.~\cite{Adler:2007ab,Adler:2007ac} are assumed). While such a rate is very small, its non-nullity entails a fundamental limitation of the theory behind the current formulation of CSL. Surely, the interaction with an external noise is expected to break energy conservation for the system alone, however, one does not expect the noise to keep transferring energy to the system forever. Thermalization to the temperature of the noise field would eventually be achieved, thus stopping the net energy increase, a mechanism that is not contemplated in the original CSL formulation. 

This has called for the proposal of a dissipative three-parameter extension (which we will dub as ``dCSL'' model)~\cite{Smirne:2014aa,Smirne:2015aa}: besides $\lcsl$ and $\rC$, the dCSL model requires the introduction of an effective temperature $T_\text{\tiny CSL}$, which can be interpreted as the temperature of the collapse noise. Dissipation guarantees that the energy of any system interacting with this noise approaches an asymptotic finite value. In the limit $\tcsl\to\infty$, which implies that the system never thermalizes with the collapse noise, one recovers  the standard CSL model, as expected. 
While there is currently no fundamental estimate of $\tcsl$, if we assume the noise to be of cosmological origin (a reasonable guess, taking into account its supposed universality), then $T_{\text{\tiny CSL}}\sim 1$\,K stands out as reasonable~\cite{Smirne:2015aa}.

The quest for the ruling-out, or the confirmation of collapse models requires the identification of a credible and physically robust framework. It is thus important to test the predictions of the dCSL model, in particular in relation to the extent to which the bounds on the CSL parameters change if one assumes a finite temperature for the collapse noise. {This analysis was initiated in the study of collapse models effects on matter-wave interferometry~\cite{Toros:2016aa} and cold atoms~\cite{Bilardello:2016aa}}. In this paper, we {extend this }investigation to optomechanical systems which now play a privileged role as they set some of the strongest bounds on the collapse parameters.

While CSL-induced effects can be easily embedded as additional noise on the motion of a mechanical system~\cite{Bahrami:2014aa}, we show that for the dCSL model this is no longer the case and  
a different strategy must be followed. Specifically, we will construct a {unitary} unravelling of the dCSL master equation, following the approach described in Ref.~\cite{Barchielli:2015aa},
which has the advantage of greatly simplifying all the necessary calculations, while providing a rigorous approach to the quantification of the effects of the collapse mechanism. Such a unitary unravelling is built around a bosonic quantum noise instead of a standard classical noise, as custom to CSL.

The paper is organized as follows: 
In Sec.~\ref{2}, after introducing the master equation for the dCSL dynamics, we build a unitary unravelling for it. In Sec.~\ref{3}, we consider a multiparticle system and 
we derive the master equation for the center of mass under the assumption of rigid body and small displacements. 
We then build a unitary unravelling for such a master equation and use it to derive the Langevin equations of motion for the center of mass of an optomechanical system, which are thus solved in Sec.~\ref{4}. The results are applied to the several experiments in Sec.~\ref{cwed} to set physically relevant bounds on the parameters characterizing the dCSL model.

\section{Unitary unravelling of the \lowercase{d}CSL model}\label{2}

The mass-proportional dCSL master equation for the density matrix $\hat \rho(t)$ of an $N$-particle system reads \cite{Smirne:2015aa}
\bq\label{dcslmaster}
\dfrac{\D \hat \rho(t)}{\D t}=-\frac{i}{\hbar}[\hat H, \hat \rho(t)]+\mathcal{L}[ \hat \rho(t)],
\eq
where $\hat H$ is the Hamiltonian of the system and
\bqali\label{Ldcslmaster}
\mathcal{L}[ \hat \rho(t)]=\nu^2\!\!\!\int\!\!\D\y\!\left(\hat L(\y)\hat\rho(t)\hat L^\dag(\y)-\tfrac12\{\hat L^\dag(\y)\hat L(\y),\hat\rho(t)\}\right)
\eqali
with $\nu=\sqrt{\lcsl \rC^3(4\pi)^{3/2}}/m_0$, where $m_0$ is the mass of each particle and $\y$ is a spatial coordinate. The Lindblad operator $\hat L(\y)$ is defined as~\footnote{Note that, in order to include the dissipation into the standard CSL model, different choices for 
$\hat L(\y)$ are possible. The one in Eq.~\eqref{defLy} resembles dissipative collisional decoherence (e.g., see~\cite{Petruccione:2005aa}), with the same interpretation for the mechanism of energy exchange with the noise~\cite{Smirne:2015aa}.}
\bqali\label{defLy}
\hat L(\y)=\frac{m_0}{(2\pi\hbar)^3}\sum_{n=1}^N\int\D\Q\,e^{\tfrac{i}{\hbar}\Q\cdot(\hat{\bf x}_n-\y)}\cdot\\
\cdot\exp\left(-\frac{\rC^2}{2\hbar^2}\left|	(1+\chi)\Q+2\chi\hat{\bf p}_n\right|^2\right),
\eqali
where $\hat \x_n$ and $\hat \p_n$ denote the position and the momentum operator of the $n$-th particle of the system, respectively, and the dimensionless parameter $\chi$ is related to the dCSL temperature $\tcsl$ by the relation
\bq\label{chie}
\chi=\frac{\hbar^2}{8m_0\kB T_{\text{\tiny CSL}}\rC^2},
\eq
where $\kB$ is the Boltzmann constant.

We wish to construct a unitary unravelling of Eq.~\eqref{dcslmaster}, i.e.~a unitary dynamics $\hat{\mathcal U}_t$ for the state vector $|\psi\rangle$  of the system such that $\hat \rho(t)=\mathbb E[\hat{\mathcal U}_t\ket{\psi}\bra{\psi}\hat{\mathcal U}^{\dagger}_t]$ is solution of the master equation. Here $\mathbb E[\ \cdot\ ]$ denotes the stochastic average over the noise. 
For the CSL model, it is straightforward to show that a classical noise is perfectly suited, as the associated Lindblad operators that can be obtained from Eq.~\eqref{defLy} by setting $\chi=0$, are self-adjoint. For the dCSL model, this is no longer possible in light of the lack of self-adjointedness of $\hat L(\y)$.  
Ref.~\cite{Hudson:1984aa} shows the way around: given a master equation in the Lindblad form [such as Eq.~\eqref{dcslmaster}], it is always possible to build a unitary unravelling by introducing quantum noise operators describing the effects of a bosonic bath. 

We thus consider the following stochastic differential equation for the state vector 
\bq\label{dU1}
\D \ket{\psi_t}=\D \mathcal U_t \ket\psi=\left\{-\frac{i}{\hbar}{\hat H}\D t+\D\hat C-\frac12\mathbb E\left[\D\hat C^\dag\D\hat C\right]\right\}\ket{\psi_t},
\eq
where $\hat C$ is a quantum noise operator that is assumed to take the following form
\bq
\hat C=\nu\int\D\y\,\left(\hat L(\y)\, \hat B^\dag(\y)-\hat L^\dag(\y)\, \hat B(\y)\right).
\eq
Here $\hat B(\y)$ is a noise field operator, whose statistical properties are identified by the It\^o rules
\bqali\label{noiserule}
\mathbb E[\D\hat B_t(\x)]=\mathbb E[\D\hat B_t^{\dag}(\x)]&=\mathbb E[\D\hat B_t^\dag(\y)\D\hat B_t(\x)]=0,\\
\mathbb E[\D\hat B_t(\y)\D\hat B_t^\dag(\x)]&=\delta(\y-\x)\,\D t.
\eqali
Eq.~\eqref{dU1} leads to a unitary evolution of the system and a simple application of It\^o rules shows that it  leads to Eq.~\eqref{dcslmaster} for the density matrix. For a more exhaustive description of stochastic Schr\"odinger equations under the action of a quantum noise, we refer to Ref.~\cite{Barchielli:2015aa}.    

\section{Master equation for the motion of the center of mass of a mechanical resonator}\label{3} 

Let us denote with ${\x}_{n}^{(0)}$ ($n=1,\dots,N$) the classical equilibrium position of each particle. We call $\mu(\x)=m_0\sum_n\delta^{(3)}(\x-\x_{n}^{(0)})$ the mass density of the system. We assume that each particle jiggles very little around its equilibrium position, so that the position operator $\hat{\bm x}_n$ of the $n$-th particle can be  written as~\cite{Nimmrichter:2014aa,Belli:2016aa}
\bq
\hat{\bm x}_n={\bm x}^{(0)}_{n}+\Delta \hat{\bm x}_{n}+\hat {\bm x},
\eq
where $\hat {\bm x}$ measures the fluctuations of the center of mass, while $\Delta \hat{\bm x}_{n}$ measures the remaining fluctuations of the $n$-th particle, which are not already included in $\hat {\bm x}$. Under the assumption of a rigid body, which will be the assumption we make from here on, the latter fluctuations are negligible. Consequently, we set $\Delta \hat{\bm x}_{n}=0$, and after tracing Eq.~\eqref{Ldcslmaster} over the relative degrees of freedom, we obtain the dissipator for the master equation for the center of mass state $ \rhocm$
\bqali\label{masteralphabeta}
\mathcal{L}[  \rhocm(t)]=\frac{\nu^2}{(2\pi\hbar)^3}\int\D\Q\,|\tilde\mu(\Q)|^2e^{-\tfrac{\rC^2(1+\chi)^2}{\hbar^2}\Q^2}\cdot\\
\cdot\left[\hat S(\Q)\rhocm(t)\hat S^\dag(\Q)-\tfrac12\acom{\hat S^\dag(\Q)\hat S(\Q)}{\rhocm(t)}\right],
\eqali
where $\tilde \mu(\Q)=\int\D\x\, \mu(\x)e^{i{\Q\cdot\x/\hbar}}$ and
\bq\label{defSalpha}
\hat  S(\Q)=e^{\tfrac{i}{\hbar}\Q\cdot\hat{\x}}
\exp\left[-2\frac{\rC^2}{\hbar^2}\left(	\chi(1+\chi)\frac{\Q\cdot\hat{\bf p}}{N}+\frac{\chi^2\hat{\bf p}^2}{N^2}\right)\right].
\eq
As the motion of the centre of mass of the rigid body is assumed to have a very small amplitude, a condition that we will shortly define quantitatively, we Taylor expand $\hat S(\Q)$. To this end, it is convenient to represent Eq.~\eqref{masteralphabeta} in the position basis. The first term in the second line becomes
\bqali\label{Gdouble}
&\bra{\x}\hat S(\Q)\rhocm(t)\hat S^\dag(\Q)\ket{\x'}=e^{-\tfrac i\hbar \Q\cdot(\x'-\x)}.
\\
&\int\D\p\int\D\p'\,e^{-\tfrac i \hbar(\x\cdot\p-\x'\cdot\p')}\braket{\p|\rhocm|\p'}\cdot\\
&\cdot \exp\left[		-\frac{2\rC^2}{\hbar^2}\left(\frac{\chi(1+\chi)}{N}\Q\cdot(\p+\p'	)+\frac{\chi^2}{N^2}(	{\p^2}+{\p'^2})	\right)	\right].
\eqali
Due to the Gaussian factor in Eq.~\eqref{masteralphabeta}, the main contribution to the integral comes from values of $\Q$ whose modulus is smaller than, or comparable to $\hbar/\rC(1+\chi)$. One can then Taylor expand Eq.~\eqref{Gdouble} under the conditions
\begin{equation}\label{cond2}
	|{\x}'-{\x}|\ll \rC(1+\chi),\quad\text{and}\quad |{\p}|,|{\p}'|\ll \frac{N\hbar}{\rC \chi}.
\end{equation}
The same procedure can be applied to the term in Eq.~\eqref{masteralphabeta} containing the anticommutator.
Then 
Eq.~\eqref{masteralphabeta} becomes
\bqali\label{linearmasteralphabeta}
&\mathcal{L}[  \rhocm(t)]=\frac{\nu^2}{(2\pi\hbar)^3}\int\D\Q\,|\tilde\mu(\Q)|^2e^{-\tfrac{\rC^2(1+\chi)^2}{\hbar^2}\Q^2}\cdot\\
&\cdot\left(\tfrac12\com{\hat K(\Q)-\hat K^\dag(\Q)+\hat M(\Q)-\hat M^\dag(\Q)}{ \rhocm(t)}+\right.\\
&\left.+\hat K(\Q)\rhocm(t)\hat K^\dag(\Q)-\tfrac12\acom{\hat K^\dag(\Q)\hat K(\Q)}{\rhocm(t)}\right),
\eqali
with 
\bqali\label{defKM}  
\hat{K}(\Q)&=-\frac{\kappa{}}{\hbar^{2}}\Q\cdot\hat{\p}+\frac{i}{\hbar}\Q\cdot\hat{\x},\\
\hat{M}{}(\Q)&=-\frac{\kappa{}^{2}}{2\hbar^{2}(1+\chi)^{2}\rC^{2}}\hat{\p}^{2}+\frac{\kappa^{2}}{2\hbar^{4}}(\Q\cdot\hat{\p})^{2}+\\
&-\frac{i}{\hbar^{3}}\kappa(\Q\cdot\hat{\x})(\Q\cdot\hat{\p})-\frac{1}{2\hbar^{2}}(\Q\cdot\hat{\x})^{2}
\eqali
and
$\kappa={2\rC^{2}\chi(1+\chi)}/{N}$. 
Now, considering the motion of a system only in one direction (say the $x$ direction), the master equation for the center of mass state becomes
\bqali\label{dqmuplmaster}
\dfrac{\D  \rhocm(t)}{\D t}=&-\frac{i}{\hbar}[ \hat H,  \rhocm(t)]-\frac{\eta}{2}\com{\hat x}{\com{\hat x}{ \rhocm(t)}}\\
&-\frac{\gamma_{\text{\tiny CSL}}^{2}}{8\eta\hbar^2}\com{\hat p}{\com{\hat p}{ \rhocm(t)}}-\frac{i\gamma_{\text{\tiny CSL}}}{2\hbar}\com{\hat x}{\acom{\hat p}{ \rhocm(t)}}, 
\eqali
with
\begin{eqnarray}
\label{etaapp1}
\eta & = &\frac{\nu^2}{(2\pi\hbar)^3\hbar^2}\int\D\Q\,|\tilde\mu(\Q)|^2e^{-\tfrac{\rC^2(1+\chi)^2}{\hbar^2}\Q^2}Q^2_x, \;\;\;\;\\
\gamma_{\text{\tiny CSL}} & = &\eta\frac{4\rC^{2}\chi(1+\chi)}{N},
\label{gammaapp}
\end{eqnarray}
where $Q_x$ denotes the $x$ component of $\Q$.
The second and third term in the right-hand side of Eq.~\eqref{dqmuplmaster} describe decoherence in position and momentum respectively,  while the last one accounts for dissipation. 

In order to write down the stochastic unravelling, it is convenient to rewrite Eq.~\eqref{dqmuplmaster} in the Lindblad form 
\bqali\label{masteragain}
\dfrac{\D  \rhocm(t)}{\D t}=&-\frac{i}{\hbar}\left[\hat H_{\text{\tiny eff}},\rhocm(t)\right]+\\
&+\eta\left(\hat{L}{\rhocm}(t)\hat{L}^{\dagger}-\frac{1}{2}\left\{ \hat{L}^{\dagger}\hat{L},{\rhocm}(t)\right\} \right),
\eqali
where 
$\hat{L} =\hat{x}+i \varkappa\hat{p}$, $\varkappa=\frac{\gamma_{\text{\tiny CSL}}}{2\eta\hbar}$ and $\hat H_{\text{\tiny eff}} = \hat H+\frac{\gamma_{\text{\tiny CSL}}}{4}\left\{ \hat{x},\hat{p}\right\}$.
As described in the previous section, the unitary unraveling is thus given by 
\bq\label{Unravel1}
\text{d}|\psi_{t}\rangle=\left\{ -\frac{i}{\hbar}\hat H_{\text{\tiny eff}}\,\text{d}t+\text{d}\hat{C}-\frac{\eta}{2}\hat{L}^{\dagger}\hat{L}\,\text{d}t\right\} |\psi_{t}\rangle  ,
\eq
where
$\text{d}\hat{C}=\hat{L}\,\text{d}\hat{B}_{t}^{\dagger}-\hat{L}^{\dagger}\,\text{d}\hat{B}_{t}$,
and the only non-zero term of the It\^o rules for the quantum noise operator is
\bq\label{noise1}
\mathbb{E}\left[\text{d}\hat{B}_{t}\text{d}\hat{B}_{t}^{\dagger}\right]=\eta\,\text{d}t.
\eq
Making use of the unravelling in Eq.~\eqref{Unravel1}, it is now rather straigtforward to derive the Langevin equations for $\hat x{}$ and $\hat p$, moving to the Heisenberg picture. 
In general, given the unitary state evolution $|\psi_t\rangle=\hat{\mathcal U}_t|\psi_0\rangle$, the stochastic variation of a generic operator $\hat O$ reads
\bqali\label{diffO00}
\D \hat O(t)=\D\hat {\mathcal U}_t^\dag\,\hat O\,\hat{\mathcal U}_t+\hat {\mathcal U}_t^\dag\,\hat O\,\D\hat{\mathcal U}_t+\mathbb E[\D\hat {\mathcal U}_t^\dag\,\hat O\,\D\hat{\mathcal U}_t],
\eqali
where the last term accounts for the It\^o contribution. 

Starting from the unravelling describing the center of mass motion in Eq.~\eqref{Unravel1}, we find the time evolution for the generic operator ${\hat O}(t)$ by differentiating Eq.~\eqref{diffO00} with respect to time (from here on, we will omit the explicit time dependence of all the operators but the noises):
\bqali\label{diffO}
\frac{\D \hat O}{\D t}&=\frac{i}{\hbar}\com{\hat H_{\text{\tiny eff}}}{\hat O}+\eta\left(\hat L^\dag\hat O\hat L-\tfrac12\acom{\hat L^\dag\hat L}{\hat O}\right)+\\
&+\left(\hat b^\dag(t)\com{\hat O}{\hat L}+\hat b (t)\com{\hat L^\dag}{\hat O}\right),
\eqali
 where we introduced $\hat b(t) =\tfrac{\D}{\D t} \hat B_t$, whose only non-zero correlation reads
\bq\label{corr:bb}
\mathbb E[ \hat b(t)\hat b^\dag(s)]=\eta\,\delta(t-s).
\eq
The corresponding Langevin equations for $\hat O=\hat x, \, \hat p$ are
\bqali \label{langevin1}
\frac{{\D \hat{x}}}{\D t}&= \frac i\hbar \com{\hat H}{\hat x}
-\varkappa  \hbar \, \hat w_x(t),\\
\frac{{\D \hat{p}}}{\D t}&=\frac i\hbar \com{\hat H}{\hat p}-\gamma_{\text{\tiny CSL}} \hat{p}  
- \hbar  \hat w_p(t),
\eqali
where we introduced $ \hat w_x(t)=\hat{b}^{\dagger }(t)+ \hat{b}(t)$ and $ \hat w_p(t)=i(\hat{b}^{\dagger }(t)- \hat{b}(t))$, whose correlations follow from Eq.~\eqref{corr:bb}
\bqali
&\mathbb E[ \hat w_x(t) \hat w_x(t')]=\mathbb E[ \hat w_p(t) \hat w_p(t')]=\eta\delta(t-t'), \\ 
&\mathbb E[ \hat w_x(t) \hat w_p(t')]=-\mathbb E[ \hat w_p(t) \hat w_x(t')]=i\eta\delta(t-t').
\eqali
Compared to the classical Langevin equation, an extra noise appears in the equation for the position operator. This is in agreement with the results in Ref.~\cite{Barchielli:2015aa}, where it is also discussed how the presence of this noise, which in this context appears naturally, is required for having a well defined momentum operator.

\section{Application to optomechanics}\label{4}

Let us consider a one-dimensional mechanical resonator of mass $m$ in an externally driven cavity. Assuming the relevant coordinate to be along the $x$ direction, the resonator and cavity field are coupled according to the radiation pressure Hamiltonian $\hat H_{\text{rp}}=\hbar g\hat a^\dag\hat a\hat x$ with $\hat a$ and $\hat a^\dag$ the annihilation and creation operators of the cavity field, $\hat x$ that should now be interpreted as the position operator for the centre of mass of the resonator, and $g$ the optomechanical coupling rate. The radiation pressure term enters the total Hamiltonian of the system, which comprises the free dynamics of the field and resonator characterized by the frequency $\omegaC$ and $\omega_0$, respectively. The motion of the system is thus described by the Langevin equations~\cite{Mancini:1994aa}
\begin{subequations}\label{free.langevin}
\begin{align}
\frac{\D{\hat x}}{\D t}&={\hat p/m},\\
\frac{\D{\hat p}}{\D t}&=-m\omega_0^2\hat x+\hbar g\hat a^\dag\hat a-\gamma_m \hat p+\hat \xi,\label{free.langevin.a}\\
\frac{\D{\hat a}}{\D t}&=-i\Delta_0 \hat a+i g\hat a\hat x-\kappa \hat a+\sqrt{2\kappa}\ain.\label{free.langevin.b}
\end{align}
\end{subequations}
The terms $-\gamma_m \hat p$ and $\hat \xi$ in Eq.~\eqref{free.langevin.a} describe the dissipative (at rate $\gamma_m$) and stochastic action of the phononic environment (at temperature $T$) affecting the mechanical resonator~\cite{Caldeira:1983aa,Hu:1992aa,Ford:2001aa,Diosi:2014aa,Ferialdi:2016aa,Carlesso:2016aa}. Here, $\hat \xi$ is an environment noise operator having zero mean and correlation function 
\bq\label{corr_env_noise}
\mathbb E[{\hat\xi_t\hat \xi_s}]
=\hbar m\gamma_m\int\frac{\D\omega}{2\pi}e^{-i\omega(t-s)}\omega\left[1+\coth\left(\tfrac{\hbar\omega}{2\kB T}\right)\right],
\eq
with $\kB$ the Boltzmann constant. In Eq.~\eqref{free.langevin.b}, $\Delta_0=\omegaC-\omega_\text{\tiny L}$ is the detuning between the cavity frequency $\omegaC$ and the frequency of the external driving field $\omega_\text{\tiny L}$. Moreover, $\kappa$ is the cavity dissipation rate and $\ain=\alpha_\text{\tiny in}+\delta\ain$ describes the driving field, characterized by the steady average amplitude $\alpha_\text{\tiny in}=\sqrt{P_\text{\tiny in}/(\hbar\omegaC)}$, where $P_\text{\tiny in}$ is the input power, and a fluctuating part that is quantum mechanically accounted for by the fluctuation operator $\delta\ain$ such that $\braket{\delta\ain(t)}=0$ and $\braket{\delta\ain(t)\delta\ain^\dag(s)}=\delta(t-s)$.

The steady-state density noise spectrum of the mechanical motion provides an informative inference tool for the long-time properties of the resonator~\cite{Gardiner:2004aa,Paternostro:2006aa}. It 
is defined as
\bqali\label{sd-def-the}
\DNS&=\frac12\int_{-\infty}^{+\infty} \D \tau \,e^{-i \omega \tau}{\mathbb E} [\braket{\{\delta\hat x(t), \delta \hat x(t+\tau)\}}],\\
&=\frac1{4\pi}\int_{-\infty}^{+\infty}\D \omega'\ \mathbb E[\braket{\{\delta \tilde x(\omega),\delta \tilde x(\omega')\}}],
\eqali
where 
$\delta \hat x(t)=\hat x(t)-\hat x_\text{\tiny st}$ is the fluctuation around the steady-state position $\hat x_\text{\tiny st}=\lim_{t\rightarrow\infty}\hat x(t)$, and $\delta\tilde x(\omega)$ denotes the Fourier transform of $\delta \hat x(t)$. 

Our goal now is to explicitly compute $\DNS$ in Eq.~\eqref{sd-def-the}, under the assumption of the dCSL dynamics for the mechanical resonator. 
To this end, we modify the set of optomechanical Langevin equations according to the prescriptions in Eq.~\eqref{langevin1}. We thus get
\begin{subequations}\label{langevin.all}
\begin{align}
\frac{\D{\hat x}}{\D t}&=\frac{\hat p}{m}-\varkappa  \hbar \, \hat w_x(t),\\
\frac{\D{\hat p}}{\D t}&=-m\omega_0^2\hat x+\hbar g\hat a^\dag\hat a-\gamma \hat p+\hat \xi
- \hbar  \hat w_p(t),\\
\frac{\D{\hat a}}{\D t}&=-i\Delta_0 \hat a+i g\hat a\hat x-\kappa \hat a+\sqrt{2\kappa}\ain,
\end{align}
\end{subequations}
where $\gamma=\gamma_m+\gamma_{\text{\tiny CSL}}$ is the total damping rate. 

We move to the frequency domain, where the equations above become algebraic, and find
\bq\label{deltaxlaser}
\delta\tilde x(\omega)=\frac{\tilde \xi(\omega)+\tilde{\mathcal N}_\text{\tiny C}(\omega)+\tilde{\mathcal N}_\text{\tiny CSL}(\omega)}{d(\omega)},
\eq
where $d(\omega)=m[(\omega_\text{\tiny eff}^2(\omega)-\omega^2)-i\gamma_\text{\tiny eff}(\omega)\omega]$ depends on the effective resonance frequency $\omega_\text{\tiny eff}(\omega)$ and damping $\gamma_\text{\tiny eff}(\omega)$, whose full expressions are given in Appendix~\ref{app.noiselaser}. Three independent sources of noise contribute to $\delta\tilde x(\omega)$: $\tilde \xi(\omega)$, which is the Fourier transform of $\hat \xi$, accounts for the phononic noise inducing Brownian motion of the mechanical system; 
$\tilde{\mathcal N}_\text{\tiny C}(\omega)$ is the source of noise due to the open nature of the cavity and induced by the driving field, and its explicit expression is given in Appendix \ref{app.noiselaser}; finally, $\tilde{\mathcal N}_\text{\tiny CSL}(\omega)$ refers to the dCSL contribution to the noise, and is the key of our analysis. It reads
\bq\label{defNcsl}
\tilde{\mathcal N}_\text{\tiny CSL}(\omega)=\varkappa\hbar m(i\omega-\gamma)\tilde w_x(\omega)-\hbar \tilde w_p(\omega),
\eq
where $\tilde w_x(\omega)$ and $\tilde w_p(\omega)$ are, respectively, the Fourier transform of $\hat w_x(t)$ and $\hat w_p(t)$. 
It is worth remarking that the dCSL noise enters $\DNS$ not only through $\tilde{\mathcal N}_\text{\tiny CSL}(\omega)$, but also in light of the presence of $\gamma_\text{\tiny CSL}$ in $d(\omega)$. The density noise spectrum of the mechanical system then reads
\begin{widetext}
\bq\label{dns1}
\DNS=\frac{1}{|d(\omega)|^2}\left[\hbar \gamma_mm\omega\coth\left(\tfrac{\hbar\omega}{2\kB T}\right)+\frac{2\hbar^2g^2\kappa^2|\alpha|^2(\Delta^2+\kappa^2+\omega^2)}{\left[\kappa^2+(\Delta-\omega)^2\right]\left[\kappa^2+(\Delta+\omega)^2\right]}	+\hbar^2\eta\left(1+\varkappa^2m^2(\gamma^2+\omega^2)\right)\right],
\eq
\end{widetext}
where $\alpha=\braket{\hat a}=\sqrt{2\kappa}\alpha_\text{\tiny in}/(\kappa-i\Delta)$ and $\Delta=\Delta_0-g\braket{\hat x}$.  
Eq.~\eqref{dns1} can be used to test the dCSL model in optomechanical experiments, to compare the corresponding predictions with those computed for the CSL model~\cite{Bahrami:2014aa,Nimmrichter:2014aa,Vinante:2016aa,Belli:2016aa,Carlesso:2016ac,Carlesso:2018ab}.

\section{Characterization of the \lowercase{d}CSL model: Comparison with experimental data}\label{cwed}

We can now apply the theoretical framework derived in the previous Sections to set experimental upper bounds on the dCSL parameters. 
We focus on nanomechanical cantilevers~\cite{Vinante:2016aa,{Vinante:2016ab}} and gravitational wave detectors~\cite{Carlesso:2016ac,Helou:2016aa}. {These are the optomechanical experiments whose data set the strongest bounds on $\lambda$ and $\rC$ for the standard CSL model}. We first perform the theoretical analysis of the setups and then we make a comparison with the experimental data.

\subsection{Nanomechanical cantilever}\label{nc}

In~\cite{Usenko:2011aa,Vinante:2016aa, Vinante:2016ab} the position variance of a cantilever, which is proportional to its temperature,  is measured for different temperatures of the surrounding environment. For our analysis, we consider the experiment reported in \cite{Vinante:2016ab}. 
The system consists of a silicon cantilever, of size $450\times57\times2.5\,\mu$m, stiffness $k_\text{\tiny stiff}=(0.40\pm0.02)\,$N/m and density 2330\,kg/m$^3$, to which a ferromagnetic micro-sphere (radius $15.5\,\mu$m and density 7430\,kg/m$^3$) is attached.
The latter has two functions: it increases the effect of the CSL noise on the system (being its density much bigger than that of silicon) and allows to monitor the motion with a SQUID in place of a laser, as considered before~\footnote{As for the laser, also the SQUID disturbs the system. This can be directly taken into account in the data extrapolation, see~\cite{Vinante:2016aa,Vinante:2016ab} for a detailed discussion.}. 
Then, without the laser contribution,  $\DNS$ becomes
\bq\label{eq.dns.cant}
\DNS=\frac1{m^2}\frac{2m\gamma_m\kB T+\hbar^2\eta\left[1+\varkappa^2m^2\left(\gamma^2+\omega^2\right)	\right]}{(\omega_0^2-\omega^2)^2+\gamma^2\omega^2},
\eq
where $\gamma=\gamma_m+\gamma_\text{\tiny CSL}$, $\omega_0=\sqrt{k_\text{\tiny stiff}/m}$ and we have considered the high temperature limit for the environmental noise. For further details we refer to \cite{Vinante:2016ab}.
By integrating $\DNS$ around the resonant frequency we obtain the temperature $T_\text{\tiny S}$ of the system
\bq
T_\text{\tiny S}=\frac{m\omega_0^2}{\kB}\int\D\omega\,\DNS=T+ \Delta T_\text{\tiny dCSL},
\eq
where $T$ is the environmental temperature and $ \Delta T_\text{\tiny dCSL}$ the dCSL contribution. The expression of the latter is given by
\bq\label{deltaTdcsl}
 \Delta T_\text{\tiny dCSL}=\frac{\hbar^2\eta\left[1+\varkappa^2m^2\left(\gamma^2+\omega_0^2\right)	\right]}{2\kB m\gamma}-\frac{\gamma_\text{\tiny CSL}}{\gamma}T.
\eq
The first term increases the temperature of the system (similarly to the standard CSL case), while the second term cools the system and this is a fingerprint of the dCSL model. To make an explicit example, if one considers an experiment where the environmental temperature is much higher than $T_\text{\tiny CSL}$, then the system is cooled by the dCSL noise, contrary to the CSL case, where the system can be only warmed up~\cite{Bilardello:2016aa}. 
\begin{figure*}[t!]
\centering 
\includegraphics[width=0.5\linewidth]{{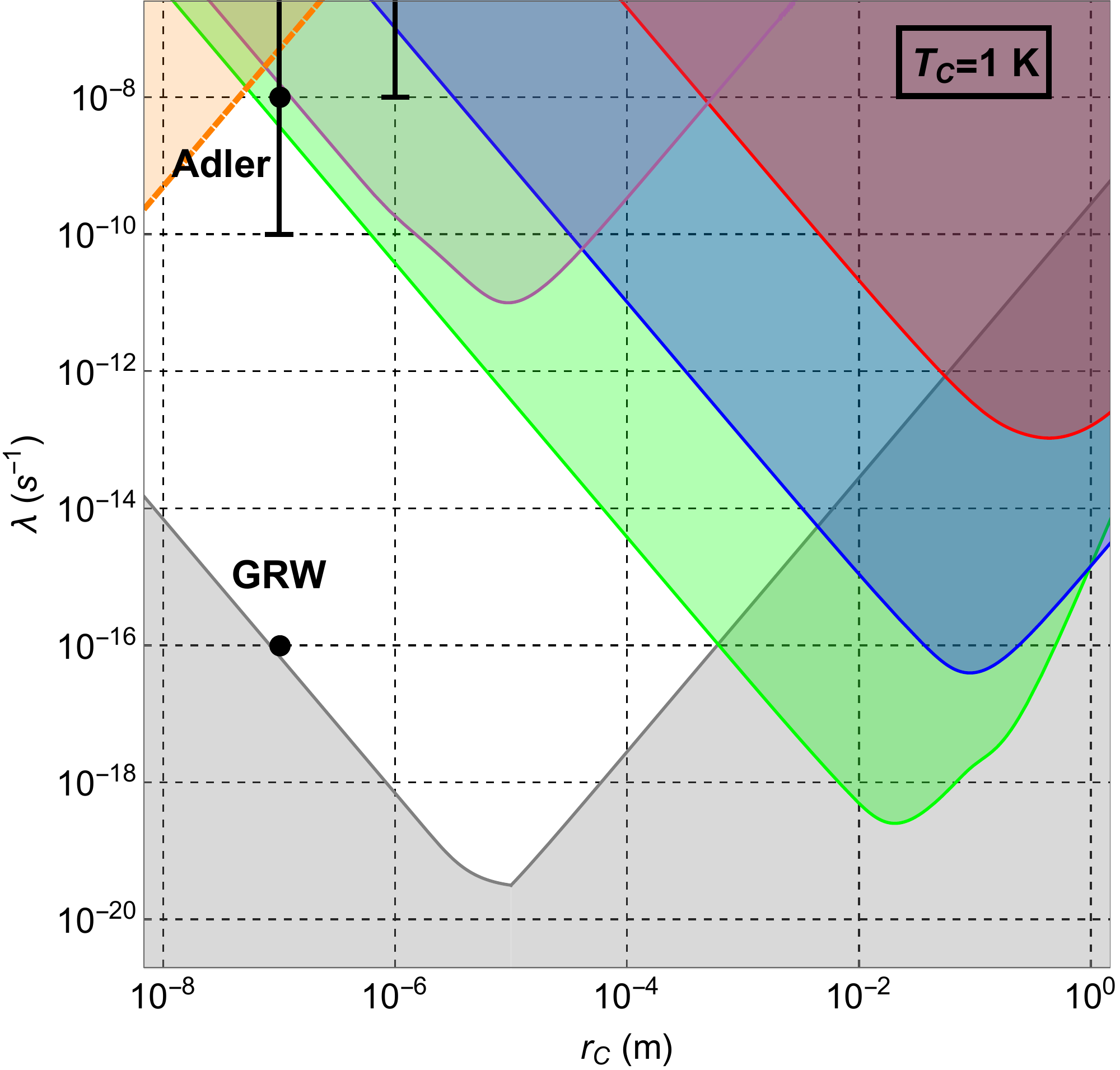}}\includegraphics[width=0.5\linewidth]{{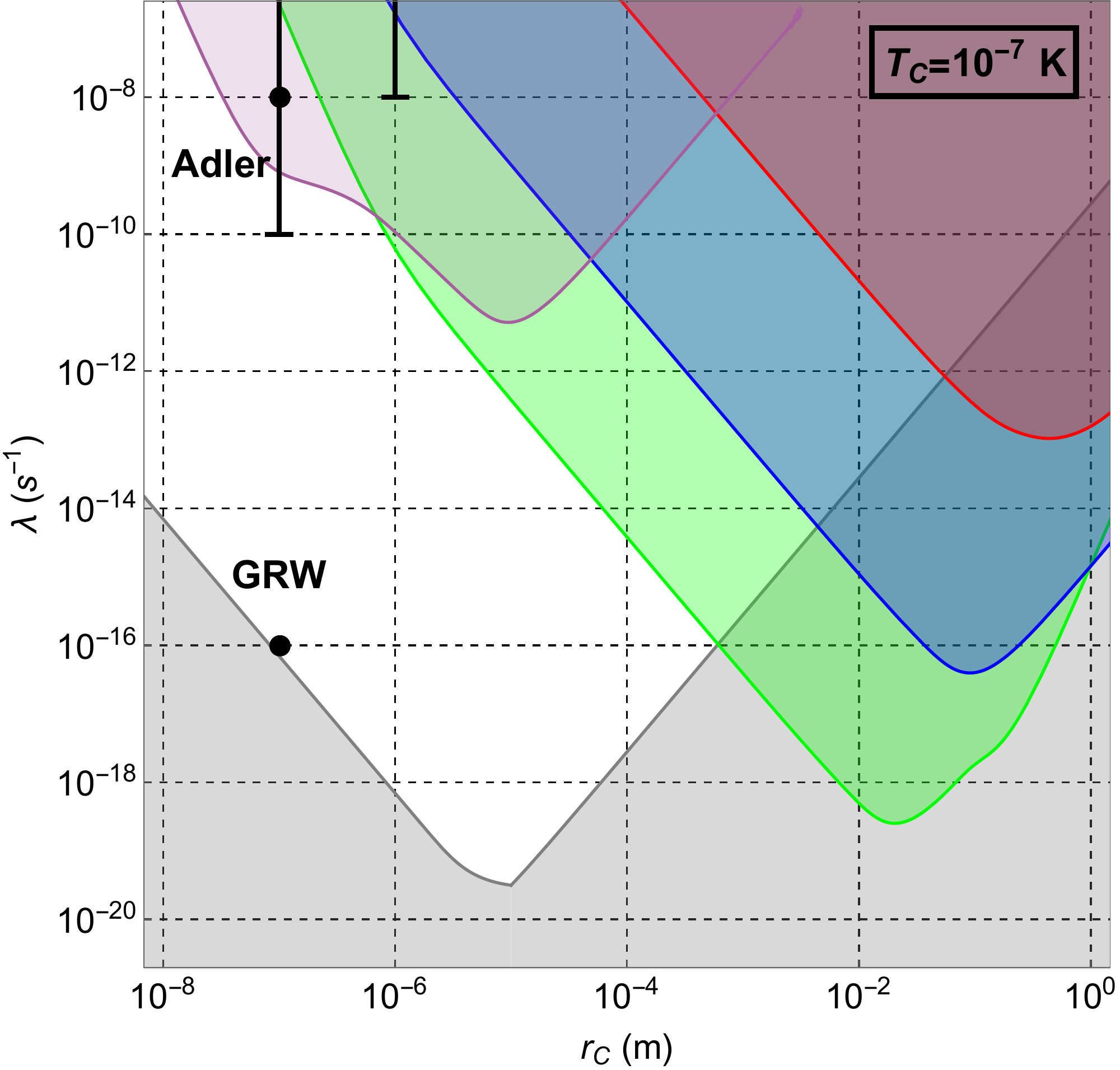}}
\caption{(Color online) {Experimental bounds on the dCSL parameters $\lambda$ and $\rC$ for {two} values of $T_\text{\tiny CSL}$. {Left panel $T_\text{\tiny CSL}=1\,$K, and Right panel $T_\text{\tiny CSL}=10^{-7}\,$K.} 
} Purple (top-center) line and shadowed area: upper bound from the cantilever experiment \cite{Vinante:2016ab}. Green, blue and red (top-right, from left to right) lines and corresponding shadowed areas: upper bounds from gravitational wave detectors, respectively LISA Pathfinder \cite{Armano:2018aa}, LIGO \cite{Abbott:2016ab} and AURIGA \cite{Vinante:2006aa}. Orange (top-left) and grey (bottom) regions: upper bound from cold atom experiment \cite{Kovachy:2015ab,Bilardello:2016aa} and lower bound from theoretical arguments \cite{Toros:2016aa}. The GRW \cite{Ghirardi:1986aa} and the Adler \cite{Adler:2007ab,Adler:2007ac} values are reported in black.
\label{fig1}}
\end{figure*}
\subsection{Gravitational wave detectors}

Following the analysis performed in \cite{Carlesso:2016ac}, we can easily derive the dCSL experimental bounds from gravitational wave detectors. The three experiments considered here are AURIGA \cite{Vinante:2006aa}, Advanced LIGO \cite{Abbott:2016ab} and LISA Pathfinder \cite{Armano:2018aa}.

AURIGA consists in an aluminium cylinder of radius $0.3\,$m, length $3\,$m and mass $2300\,$kg cooled at 4.2\,K, whose resonant deformation at frequency $\omega_0/2\pi\sim900\,$Hz is monitored by a SQUID-based readout \cite{Baggio:2005aa}. We model the system with two cylinders of half length, oscillating in counterphase, as done in \cite{Carlesso:2016ac}. The minimum value for the force noise, which could be attributed to dCSL \cite{Carlesso:2016ac}, is $\mathcal S_\text{\tiny F}=12\,$pN/Hz$^{1/2}$.

LIGO is a Michelson interferometer, whose two arms are configured as a Fabry-Perot cavity, with two cylindrical silica mirrors (density 2200\,kg/m$^3$, radius 17\,cm and length 20\,cm)  separated by a distance of 4\,km. We estimate that the minimum effective noise $\mathcal S_\text{\tiny F}=95\,$fN/Hz$^{1/2}$ is reached at $\omega/2\pi=30-35$\,Hz  \cite{Abbott:2016ab,Abbott:2016aa}.

LISA Pathfinder consists in a pair of cubical masses (mass 1.928\,kg and side length 4.6\,cm) which are  37.6\,cm away from each other. The two masses are in free fall, surrounded by a space satellite following them, and orbiting around the first Lagrangian point of the Sun-Earth system. The minimum force noise is $\mathcal S_\text{\tiny F}=1.77\,$fN/Hz$^{1/2}$ just above mHz regime \cite{Armano:2018aa}.

Differently from the cantilever, where one measures the center-of-mass motion, here the relevant quantity is the relative distance $\R_{12}$ between the two masses (in the case of AURIGA this corresponds to the elongation of the single mass).
Then, the equations of motion must be changed accordingly. We explicitly derive them in Appendix \ref{app:composite} and we obtain for the corresponding $\DNS$
\bq\label{dns2A}
\DNS=\frac{\hbar^2(\eta-\sigma)}{m^2}\,\frac{1+m^2 \varkappa^2(\gamma^2+\omega^2)}{(\tilde\omega_0^2-\omega^2)^2+\tilde\gamma^2\omega^2},
\eq
where $\tilde \omega_0^2=\omega_0^2-2\gamma \varkappa \sigma\hbar$, $\tilde\gamma=\gamma -2 \varkappa\sigma \hbar$ and the explicit form of $\sigma$ is given in Eq.~\eqref{sigmaA}. Since we are primarily interested in estimating the effect of the dCSL noise, we neglect all other noise sources, paying the price of setting more conservative bounds.

\subsection{Bounds on dCSL parameters}\label{subsec.exp}

In Fig.~\ref{fig1} we report the bounds on the parameters $\lcsl$ and $\rC$ by choosing two different values of $T_{\text{\tiny CSL}}$. The value of $T_{\text{\tiny CSL}}=1\,$K is a natural choice if one assumes that the CSL noise has a cosmological origin. Compared to the results presented in \cite{Bilardello:2016aa,Carlesso:2016ac,Carlesso:2018ab,Vinante:2016ab}, which refer to the CSL model ($T_{\text{\tiny CSL}}=+\infty$), the first panel shows no appreciable difference. Hence, for any $T_{\text{\tiny CSL}} >1\,$K bounds on the dCSL model are practically equivalent to those on the standard CSL model.
 
{ Things start changing} if we take different values for the noise temperature. Specifically, we consider as an example the value of {$T_\text{\tiny CSL}=10^{-7}\,$K}.
As Fig.~\ref{fig1} shows, the bounds from gravitational wave detectors are stable, still coinciding with those obtained in \cite{Carlesso:2016ac,Helou:2016aa,Carlesso:2018ab} with the reference to CSL model. The reason is that the diffusion constant $\eta$ defined in Eq.~\eqref{etaapp1} is the only relevant quantity here, and it changes with respect to the CSL model only if $1+\chi$ cannot be approximated to unity. This takes place for ranges of the noise temperature such that [cf.~Eq.~\eqref{chie}]
\bq
T_{\text{\tiny CSL}}\rC^2\ll\frac{\hbar^2}{8m_0\kB }\sim10^{-18}\,\text{m$^2$K}.
\eq
Thus, changes are expected for $T_\text{\tiny CSL}\leq10^{8}\,$K when $\rC\ll10^{-13}\,$m and for $T_\text{\tiny CSL}\leq10^{-7}\,$K when $\rC\ll10^{-5}\,$m. This can be seen in the bound coming from LISA Pathfinder, which becomes slightly weaker for $\rC<10^{-6}\,$m at $T_\text{\tiny CSL}=10^{-7}\,$K as shown in the right panel of Fig.~\ref{fig1}. 

A strong effect of the dissipative extension of the model is shown in the bounds from the nanomechanical cantilever for $T_\text{\tiny CSL}=10^{-7}\,$K. Such a change is driven not only by changes in $\eta$ as discussed before, but also by the change of the dissipation rate $\gamma=\gamma_m+\gamma_\text{\tiny CSL}$, with $\gamma_\text{\tiny CSL}=\eta\gamma'$, where from Eq.~\eqref{gammaapp} we have
\bq
\gamma'= \frac{4\rC^{2}m_0\chi(1+\chi)}{m}.
\eq
Moreover, for this experiment there is an additional dCSL contribution which, conversely to the case of the gravitational wave detectors experiments considered before, is not negligible. This comes from the 
 term $\varkappa^2m^2(\gamma^2+\omega^2)$ in $\DNS$ as defined in Eq.~\eqref{dns1}, where 
\bq\label{varkappam}
\varkappa m=\frac{\hbar}{4\kB T_{\text{\tiny CSL}}}\left(1+\frac{\hbar^2}{8m_0\kB T_{\text{\tiny CSL}}\rC^2}\right)
\eq
is independent from the system parameters. The term $\varkappa^2m^2(\gamma^2+\omega^2)$ becomes relevant when significantly larger than 1. For the cantilever under consideration, the transition occurs at $T_\text{\tiny CSL}\lesssim 10^{-5}\,$K for $\rC=10^{-8}\,$m and at $T_\text{\tiny CSL}\lesssim 10^{-7}\,$K for $\rC=10^{-4}\,$m.
 Such a term affects the system more than the modification of the diffusion constant, and consequently the corresponding bound becomes stronger for small $\rC$.

\section{Conclusions}

We have provided a description of the dCSL model in terms of Langevin equations resulting from a unitary unravelling of the collapse master equation. Our linear and unitary unravelling is able to mimic the non-linear and stochastic action of the dCSL model, including its dissipative nature. The approach that we have put forward is fully suited for optomechanical setups such as cantilevers and  gravitational wave detectors, which were discussed in Sec.~\ref{cwed}.

We have identified the bounds on dCSL parameters $\lcsl$ and $\rC$ for two values of the noise temperature $T_\text{\tiny CSL}$. For $T_{\text{\tiny CSL}}>1\,$K the dissipative effects are negligible and the bounds are {\it de facto} the same as those obtained with the standard CSL model \cite{Carlesso:2016ac,Carlesso:2018ab}. For $T_{\text{\tiny CSL}}=10^{-7}\,$K, the cantilever bound in the region $\rC\ll 10^{-5}\,$m is modified. Conversely, the bounds given by gravitational wave detectors are almost completely unaffected by such a dissipative extension for the considered ranges of temperatures. Lowers values of the temperature seem unrealistic and therefore were not considered.

Our approach can be in suitably applied also to other non-interferometric tests of collapse models, such as spontaneous photon emission from Germanium \cite{Fu:1997aa,Curceanu:2016aa,Piscicchia:2017aa} and phonon excitations in crystals \cite{Adler:2018aa,Bahrami:2018aa}. However, in this case the conditions in Eq.~\eqref{cond2} are not fulfilled, therefore the approximations used through the text cannot be applied and one has to proceed in a different way. One should note that these bounds, coming from photon emission and phonon excitations, significantly depend on the spectrum of the noise and disappear for a frequency cut-off in the range $10^{11}-10^{15}\,$Hz \cite{Adler:2007ad,Adler:2013aa,Donadi:2014aa,Bassi:2014aa,Donadi:2015aa,Carlesso:2018aa}. Therefore, analyzing how these bounds are affected by dissipative effects seems not so relevant.

Our investigation is well placed within the current research effort towards the sharpening of collapse models in light of possible (and indeed foreseeable) experimental assessment of their effects on massive systems. We believe that {\it curing} a physically significant drawback of CSL-like mechanisms such as their inherent energy non-conserving nature provides more robust theoretical models to be contrasted to the evidence of experimental data gathered in any of the settings that we have analyzed here, and thus a more compelling case for the exploration of possible alternative models for the quantum-to-classical transition.

\section*{Acknowledgments}
We are grateful to Giulio Gasbarri and Luca Ferialdi for many useful and valuable comments on the paper. JN and AB acknowledge financial support from University of Trieste (Grant FRA 2016). MC, MP and AB acknowledge financial support the EU Collaborative Project TEQ (Grant Agreement 766900).
AB acknowledges financial support from the Instituto Nazionale di Fisica Nucleare (INFN). SD, MP and AB acknowledge COST Action CA15220 QTSpace. SD acknowledges financial support from Fondazione Angelo Della Riccia and The Foundation BLANCEFLOR Boncompagni Ludovisi, n\'ee Bildt.

\appendix

\section{Study of the validity of the conditions in Eq.~\eqref{cond2} for the analysis in section~\ref{nc}}\label{app:ce}
\begin{figure}[b!]	
	\includegraphics[width=\linewidth]{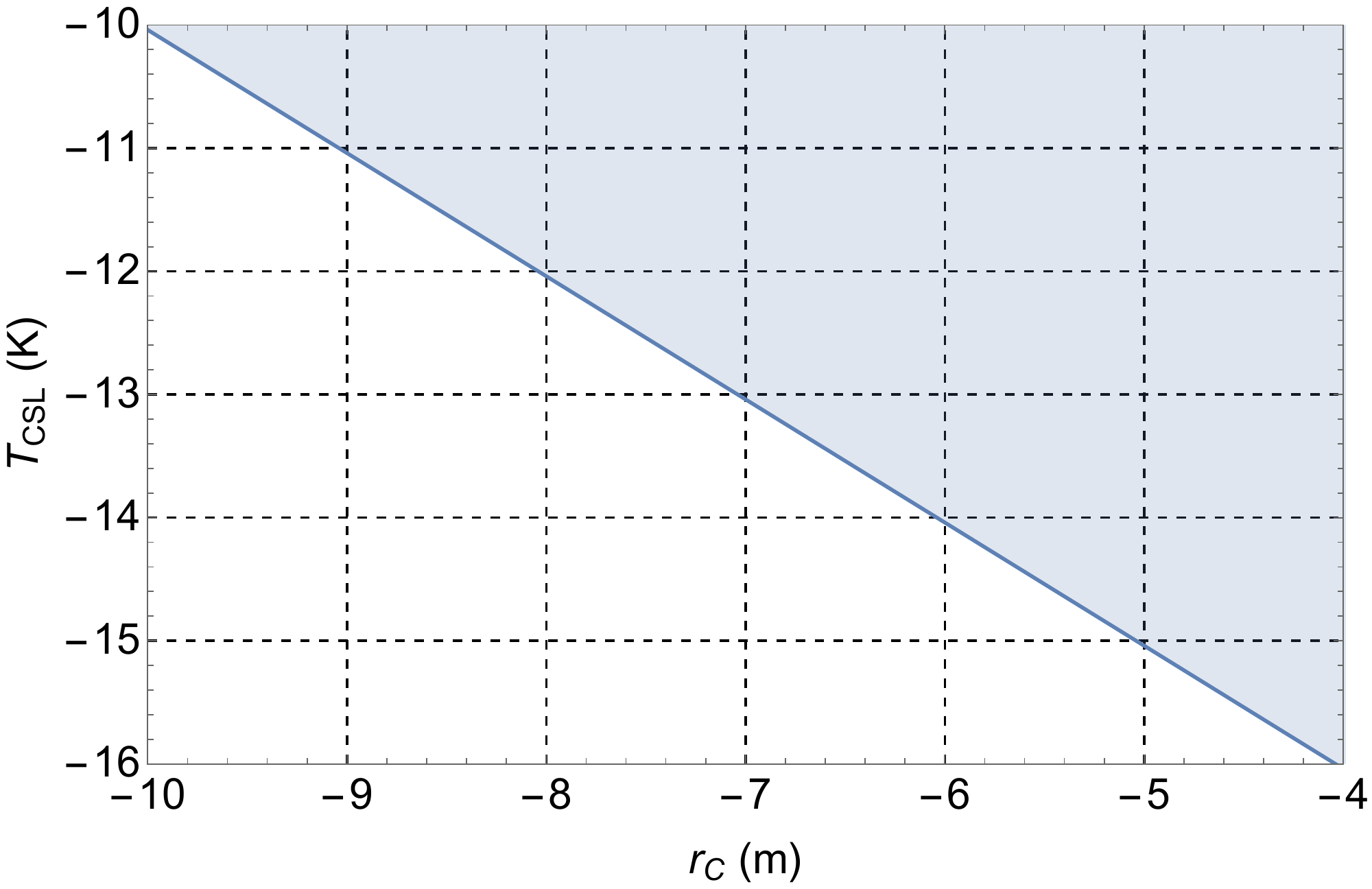}
	\caption{Regime of validity of the conditions in (Eq.~\eqref{cond2}) for the cantilever considered in section~\ref{nc}. The blue area denotes the values of $\tcsl$ and $\rC$ for which these conditions are fulfilled. }
	\label{EC}
\end{figure}
As shown in the main text, Eq.~\eqref{masteralphabeta} can be approximated by Eq.~\eqref{dqmuplmaster} when two assumptions are fulfilled, see Eq.~\eqref{cond2}, which are here reported:
\begin{equation}
	|{\x}'-{\x}|\ll \rC(1+\chi),\quad\text{and}\quad |{\p}|,|{\p}'|\ll \frac{N\hbar}{\rC \chi}.
\end{equation}
Since we work in a reference frame where the average velocity of the system is zero, a good estimation of $|\bm{v}|$ is given by its fluctuations $\Delta\bm{v}$.
We are considering the  system in the steady state i.e.~when it has thermalized with the environment, and this allows to use the equipartition theorem to estimate the fluctuations of the position and the velocity of the system:
\bq
\Delta{\x} \sim \sqrt{\frac{\kB T}{m\, \omega_{0}^{2}}},\quad\Delta\bm{v} \sim \sqrt{\frac{\kB T}{m}}.
\eq
For the cantilever considered in section~\ref{nc}, we find
\bq
\Delta{\x} \sim10^{-12}\,\text{m},\quad\Delta\bm{v} \sim10^{-8}\,\text{m/s}.
\eq
The shaded region in Fig.~\ref{EC} shows the values of $\rC$ and $\tcsl$ which fulfil the conditions of Eq.~\eqref{cond2}, taking the environmental temperature as $T \simeq 1\,$mK. As we can see, even for very low CSL temperatures, such as $\tcsl\sim10^{-10}\,$ K, for any $\rC\geq10^{-10}\,\text{m}$ conditions of Eq.~\eqref{cond2} are satisfied and the analysis in the main text is valid. Only when reaching much lower temperatures as $\tcsl\sim10^{-15}\,\text{K}$, the range of values of $\rC$ which satisfy Eq.~\eqref{cond2} strongly reduces.

\section{Density Noise Spectrum details}
\label{app.noiselaser}

The explicit form of the effective resonant frequency $\omega_\text{\tiny eff}(\omega)$, of the effective damping $\gamma_\text{\tiny eff}(\omega)$ and of the laser noise $\tilde{\mathcal N}_\text{\tiny C}(\omega)$ appearing in Eq.~\eqref{deltaxlaser} can be derived by following the standard procedure \cite{Gardiner:2004aa,Paternostro:2006aa}. Respectively, they read
\bqali
\omega_\text{\tiny eff}^2(\omega)&=\omega_0^2-\frac{2|\alpha|^2\hbar g^2\Delta(\Delta^2-\omega^2+\kappa^2)}{m\left((\Delta+\omega)^2+\kappa^2\right)(\left(\Delta-\omega)^2+\kappa^2\right)},\\
\gamma_\text{\tiny eff}(\omega)&=\gamma+\frac{4|\alpha|^2\hbar g^2\kappa\Delta}{m\left((\Delta+\omega)^2+\kappa^2\right)(\left(\Delta-\omega)^2+\kappa^2\right)},\\
\tilde{\mathcal N}_\text{\tiny C}(\omega)&=\hbar g\sqrt{2\kappa}\left[\frac{\alpha^*\tilde a_\text{\tiny in}(\omega)}{\kappa+i(\Delta-\omega)}+\frac{\alpha\tilde a^\dag_\text{\tiny in}(-\omega)}{\kappa-i(\Delta+\omega)}	\right],
\eqali
where $\Delta=\Delta_0-g\braket{\hat x}$ and  the only non zero correlation is given by $\braket{\tilde a_\text{\tiny in}(\omega)\tilde a_\text{\tiny in}^\dag(-\bar\omega)}=2\pi\delta(\omega+\bar\omega)$.

\section{dCSL for composite systems}\label{app:composite}

We consider a system containing $N$ particles, which can be divided in 2 subsets labeled by the indeces $\alpha,\beta=1,2$ where the $\alpha$-th subset has $N_\alpha$ particles. The mass density of each subset is described by $\mu_\alpha(\x)=m_0\sum_n\delta^{(3)}(\x-\x_{n,\alpha}^{(0)})$, where ${\bf x}_{n,\alpha}^{(0)}$ denotes the classical equilibrium position of the $n$-th nucleon (belonging to the $\alpha$-th mass distribution). Then, similarly to the procedure shown in the main text, we can express Eq.~\eqref{Ldcslmaster} as:
\bqali\label{masteralphabetaA}
\mathcal{L}[ \hat \rho(t)]=\frac{\nu^2}{(2\pi\hbar)^3}\sum_{\alpha,\beta}\int\D\Q\,\tilde\mu_\alpha(\Q)\tilde\mu_\beta^*(\Q)e^{-\tfrac{\rC^2(1+\chi)^2}{\hbar^2}\Q^2}\cdot\\
\cdot\left[\hat S_\alpha(\Q)\hat\rho(t)\hat S_\beta^\dag(\Q)-\tfrac12\acom{\hat S_\beta^\dag(\Q)\hat S_\alpha(\Q)}{\hat\rho(t)}\right],
\eqali
where $\tilde\mu_\alpha(\Q)=\int\D\x\,\mu_\alpha(\x)e^{i\Q\cdot\x/\hbar}$ and $\hat  S_\alpha(\Q)$ takes the same expression of $\hat  S(\Q)$ in Eq.~\eqref{defSalpha}, with the following substitutions: 
\bq\label{xtoxa}
\hat\x\to\hat\x_\alpha,\quad \hat\p\to\hat\p_\alpha \quad\text{and}\quad N\to N_\alpha.
\eq
The dissipator in Eq.~\eqref{masteralphabetaA} describes the $N$ particles system when this is considered as divided in subsets labeled by $\alpha$. 

Under the short-lenght approximation, valid for 
\begin{equation}\label{condappendix}
	|{\x}'_\beta-{\x}_\alpha|\ll \rC(1+\chi),\quad\text{and}\quad |{\p}_\alpha|,|{\p}_\beta|\ll \frac{N\hbar}{\rC \chi},
\end{equation}
Eq.~\eqref{masteralphabetaA} can be approximated with 
\bqali\label{linearmasteralphabetaA}
&\mathcal{L}[ \hat \rho(t)]=\frac{\lcsl \rC^3}{\pi^{3/2}m_0^2\hbar^3}\sum_{\alpha,\beta}\int\D\Q\,\tilde\mu_\alpha(\Q)\tilde\mu_\beta^*(\Q)e^{-\tfrac{\rC^2(1+\chi)^2}{\hbar^2}\Q^2}\\
&\cdot\left(\tfrac12\com{\hat K_\alpha(\Q)-\hat K_\beta^\dag(\Q)+\hat M_\alpha(\Q)-\hat M_\beta^\dag(\Q)}{\hat \rho(t)}+\right.\\
&\left.+\hat K_\alpha(\Q)\hat\rho(t)\hat K_\beta^\dag(\Q)-\tfrac12\acom{\hat K_\beta^\dag(\Q)\hat K_\alpha(\Q)}{\hat\rho(t)}\right),
\eqali
where $\hat K_\alpha$ and $\hat M_\alpha$ can be obtained from Eq.~\eqref{defKM} with the replacements in Eq.~\eqref{xtoxa}. Note that the first condition in Eq.~\eqref{condappendix} is fulfilled, assuming that $|\x_\alpha|$ and $|\x_\beta|\ll \rC$, even when $\alpha$ and $\beta$ belong to different subsets, centred around points distant more than $\rC$. Indeed, $\x_{\alpha}$ and $\x_{\beta}$ describe the fluctuations of the $\alpha$ and $\beta$ subsets around the corresponding centers of mass and not their actual positions. Consequently, we have $|{\x}'_\beta-{\x}_\alpha|\leq|\x_\alpha|+|\x_\beta|$, which is smaller than $\rC$ [cf.~Eq.~\eqref{cond2}].

As already stated in the main text, there are two situations of interest. The first one is when the system is not divided in subsets, i.e.~when $\alpha=\beta=1$. Then Eq.~\eqref{linearmasteralphabetaA} reduces simply to Eq.~\eqref{linearmasteralphabeta} describing the motion of the center-of-mass of the system. This first case is discussed in the main text and examples of systems which can be well described just by studying the center of mass motion are cantilevers \cite{Vinante:2016aa,Vinante:2016ab} or optical levitated nanospheres \cite{Jain:2016aa}. On the other hand, in interferometric experiments involving two masses, as LIGO and LISA Pathfinder \cite{Abbott:2016ab,Armano:2016aa}, one is interested in the relative motion between two distinct objects. In such a case the dynamics is described by Eq.~\eqref{linearmasteralphabetaA} with $\alpha,\beta=1,2$.

We now restrict to the case of two subsets having the same mass density distribution, at positions displaced by $\R_{12}$. Accordingly, $N_2=N_1$ and
\bq\label{assumptionidenticalA}
\tilde \mu_2(\Q)=\tilde \mu_1(\Q)e^{-i\Q\cdot\R_{12}/\hbar}.
\eq
Under this assumption, Eq.~\eqref{linearmasteralphabetaA} becomes:
\bqali\label{master2A}
\mathcal{L}[ \hat \rho(t)]&=\frac{\lcsl \rC^3}{\pi^{3/2}m_0^2\hbar^3}\int\D\Q\,|\tilde\mu_1(\Q)|^2e^{-\tfrac{\rC^2(1+\chi)^2}{\hbar^2}\Q^2}\cdot\\
&\cdot\left(\hat f_{11}+\hat f_{22}+\hat f_{12}e^{-i\Q\cdot\R_{12}/\hbar}+\hat f_{21}e^{i\Q\cdot\R_{12}/\hbar}\right),
\eqali
where we introduced
\bqali
\hat f_{\alpha\beta}&=\tfrac12\com{\hat K_\alpha-\hat K_\beta^\dag+\hat M_\alpha-\hat M_\beta^\dag}{\hat \rho(t)}+\\
&+\hat K_\alpha(\Q)\hat\rho(t)\hat K_\beta^\dag(\Q)-\tfrac12\acom{\hat K_\beta^\dag(\Q)\hat K_\alpha(\Q)}{\hat\rho(t)}.
\eqali
The meaning of the four terms in Eq.~\eqref{master2A} is the following: the terms $\hat f_{11}$ and $\hat f_{22}$ give, respectively, the contribution to the master equation due to the mass distributions $\mu_1$ and $\mu_2$ as if they were alone; this is the incoherent contribution. The last two terms instead account for correlation effects between the two mass distributions.

To better understand the meaning of Eq.~\eqref{master2A}, let us consider two limiting cases.
 The first limit is given by $|\R_{12}|\gg \rC$, for which the phases multiplying $\hat f_{12}$ and $\hat f_{21}$ oscillate very rapidly, giving a negligible contribution compared to that of $\hat f_{11}$ and $\hat f_{22}$. This means that for large distances the noise acting on the fist mass distribution is totally uncorrelated from the one acting on the second mass. 
In the opposite limit, i.e.~when $|\R_{12}|\ll \rC$, we have $e^{i\Q\cdot\R_{12}/\hbar}\simeq 1$. In this case, the same noise acts on the two mass distributions, and the contributions from the cross terms become relevant.

By considering the problem only along the direction of motion ($x$-direction), the master equation becomes
\bqali
\frac{\text{d}\hat\rho(t)}{\text{d}t}&=-\frac{i}{\hbar}\left[\hat{H}_{\text{\tiny eff}}^{(2)},\hat \rho(t)\right]+\\
&+\sum_{\alpha=1}^{2}K_{\alpha,\beta}\left(\hat{L}_{\alpha}\hat{\rho}(t)\hat{L}_{\beta}^{\dagger}-\frac{1}{2}\left\{ \hat{L}_{\beta}^{\dagger}\hat{L}_{\alpha},\hat{\rho}(t)\right\} \right),
\eqali
where
\bqali\label{Heff2A}
\hat{H}_{\text{\tiny eff}}^{(2)}&=\hat{H}+\left(\frac{\gamma_{\text{\tiny CSL}}}{4}+\frac{\varkappa\sigma\hbar}{2}\right)\left(\left\{ \hat{x}_{1},\hat{p}_{1}\right\}+\left\{ \hat{x}_{2},\hat{p}_{2}\right\} \right)+\\
&+\varkappa\Omega\hbar^{2}(\hat{p}_{1}-\hat{p}_{2}),
\eqali
\bq\label{sigmaA}
\sigma\!=\!\frac{\nu^2}{(2\pi\hbar)^3\hbar^2}\!\!\int \!\! \text{d}{\Q}\,|\tilde{\mu}({\Q})|^{2}e^{-\tfrac{\rC^{2}(1+\chi)^{2}}{\hbar^{2}}{\Q}^{2}}\!\!Q_{x}^{2}\cos\!\left(\frac{Q_{x}R_{12}}{\hbar}\right)\!\!,
\eq
\bq\label{BigsigmaA}
\Omega\!=\!\!\frac{\nu^2}{(2\pi\hbar)^3\hbar^2}\!\!\int\!\!\text{d}{\Q}\,|\tilde{\mu}({\Q})|^{2}e^{-\tfrac{\rC^{2}(1+\chi)^{2}}{\hbar^{2}}{\Q}^{2}}\!\!Q_{x}\sin\!\left(\frac{Q_{x}R_{12}}{\hbar}\right)\!\!,
\eq
and 
\begin{align}\label{BigKA}
&K_{\alpha,\beta}=\left(\begin{array}{cc}
\eta & \sigma\\
\sigma & \eta
\end{array}\right),\quad\text{}\quad \varkappa=\frac{\gamma_{\text{\tiny CSL}}}{2\eta\hbar},\\
\label{Lalpha2A}
&\hat{L}_{\alpha}=\hat{x}_{\alpha}+i \varkappa\hat{p}_{\alpha}\;\;\;\;\text{with}\;\;\;\;\alpha=1,2.
\end{align}
Note that the parameters $\sigma$ and $\Omega$, similarly to $\eta$ defined in Eq.~\eqref{etaapp1}, depend on the phenomenological constants $\rC$, $\lambda_{\tiny \text{CSL}}$ and  $T_{\tiny \text{CSL}}$ of the dCSL model as well as on the mass distribution of the system. In the limit when the center of mass of the two sub-systems coincide, i.e. $R_{12}\to0$, one finds that $\sigma\to\eta$ and $\Omega\to0$.

Following the same scheme of the main text, we can write the corresponding unitary unravelling:
\bq\label{unravel2A}
\text{d}|\psi_{t}\rangle=\left\{ -\frac{i}{\hbar}\hat{H}_{\text{\tiny eff}}^{(2)}\text{d}t+\text{d}\hat{C}_{2}-\frac{1}{2}\mathbb{E}\left[\text{d}\hat{C}_{2}^{\dagger}\text{d}\hat{C}_{2}\right]\right\} |\psi_{t}\rangle,
\eq
where 
\bq\label{C2A}
\text{d}\hat{C}_{2}=\sum_{\alpha=1}^{2}\left(\hat{L}_{\alpha}\,\text{d}\hat{B}_{\alpha t}^{\dagger}-\hat{L}_{\alpha}^{\dagger}\,\text{d}\hat{B}_{\alpha t}\right),
\eq
and with the It\^o rules 
\bq\label{corrnoise2A}
\mathbb{E}\left[\text{d}\hat{B}_{\alpha t}\text{d}\hat{B}_{\beta t}^{\dagger}\right]=K_{\beta,\alpha}\text{d}t,
\eq
and all the others It\^o products are zero.

Given the Eq.~\eqref{unravel2A} the Langevin equation for a generic operator 
$\hat O$ is:  
\bqali\label{diffO2A}
\frac{\D \hat O}{\D t}&=\frac{i}{\hbar}\com{\hat{H}_{\text{\tiny eff}}^{(2)}}{\hat O}+\!\sum_{\alpha=1}^{2}\left(\hat b_\alpha^\dag(t)\!\com{\hat O}{\hat L_\alpha}\!+\!\hat b_\alpha(t)\com{\hat L_\alpha^\dag}{\hat O}\right)\!+\\
&+\!\sum_{\alpha,\beta=1}^{2}\!K_{\alpha,\beta}\left(\hat L_\alpha^\dag\hat O\hat L_\beta-\tfrac12\acom{\hat L_\alpha^\dag\hat L_\beta}{\hat O}\right),
\eqali
where we introduced $\hat b_\alpha (t) =\tfrac{\D}{\D t} \hat B_{\alpha,t}$.
By considering
\bq
\hat H=\sum_{\alpha=1}^2\frac{\hat p_\alpha^2}{2m}+\frac12m\omega_0^2\hat x_\alpha^2,
\eq
in Eq.~\eqref{Heff2A},
the Langevin equations for the relative coordinates $\hat x =\hat x_1-\hat x_2$ and $\hat p =\tfrac12(\hat p_1-\hat p_2)$ of the two masses become
\bqali\label{langevin2A}
\frac{\D \hat x}{\D t}&=\frac{2\hat p}{m}+2 \varkappa\sigma\hbar \hat x+2 \varkappa\Omega\hbar^2- \varkappa \hbar \hat w_{x}(t),\\
\frac{\D \hat p}{\D t}&=-\frac{m}{2} \omega_0^2\hat x-\gamma \hat p-\frac{\hbar}{2} \hat w_{p}(t),
\eqali
where we introduced 
\bqali\label{wrel}
\hat w_x(t)&=\left[\hat{b}^{\dagger }_1(t)+ \hat{b}_1(t)\right]-\left[\hat{b}^{\dagger }_2(t)+ \hat{b}_2(t)\right],\\
\hat w_p(t)&=i\left[\hat{b}^{\dagger }_1(t)-\hat{b}_1(t)\right]-i\left[\hat{b}^{\dagger }_2(t)- \hat{b}_2(t)\right],
\eqali
with
\bqali\label{corrrel}
\mathbb E[  \hat w_{x}(t)  \hat w_{x}(s)]&=2(\eta-\sigma)\delta(t-s),\\
\mathbb E[  \hat w_{x}(t) \hat w_{p}(s)]&=2i(\eta-\sigma)\delta(t-s),\\
\mathbb E[ \hat w_{p}(t) \hat w_{x}(s)]&=-2i(\eta-\sigma)\delta(t-s),\\
\mathbb E[ \hat w_{p}(t)  \hat w_{p}(s)]&=2(\eta-\sigma)\delta(t-s),
\eqali
describing the correlations between the noises.

We now compute the density noise spectrum for the relative position. Starting from Eqs.~\eqref{langevin2A} the fluctuation in position in Fourier space is
\bq
\delta\tilde x(\omega)=\frac{\hbar}m \frac{m a(i \omega-\gamma)\tilde w_{x}-\tilde w_p}{(\omega_0^2-\omega^2-2\gamma \varkappa\sigma\hbar)-i\omega(\gamma-2 \varkappa\sigma\hbar)},
\eq
where the correlations of the Fourier transformed noises read:
\bqali
\mathbb E[ \tilde w_x(\omega) \tilde w_x(\omega')]&=4\pi(\eta-\sigma)\delta(\omega+\omega'),\\
\mathbb E[ \tilde w_x(\omega) \tilde w_p(\omega')]&=4\pi i(\eta-\sigma)\delta(\omega+\omega'), \\
\mathbb E[ \tilde w_p(\omega) \tilde w_x(\omega')]&=-4\pi i(\eta-\sigma)\delta(\omega+\omega'), \\
\mathbb E[ \tilde w_p(\omega) \tilde w_p(\omega')]&=4\pi(\eta-\sigma)\delta(\omega+\omega').
\eqali
The corresponding density noise spectrum, calculated using Eqs.~\eqref{sd-def-the} with the relative coordinates, is given in Eq.~\eqref{dns2A}.

\clearpage

\end{document}